\documentclass[12pt]{iopart} 
\usepackage{epsf}
\usepackage{epsfig} 
\usepackage{latexsym} 
\usepackage{iopams}
\usepackage{amsthm}
\def\eqref#1{(\ref{#1})}

\newcommand{\calD}{{\cal{D}}}

\newcommand{\calG}{{\cal{G}}}

\newcommand{\calL}{{\cal{L}}}

\newcommand{\calP}{{\cal{P}}}

\newcommand{\calV}{{\cal{V}}} 
\newcommand{\calZ}{{\cal{Z}}} 
\newcommand{\bbR}{{\mathbb{R}}} 
\newcommand{\bx}{{\bi{x}}} 
\newcommand{\bz}{{\bi{z}}} 
\newcommand{\by}{{\bi{y}}} 
\newcommand{\ba}{{\bi{a}}} 
\newcommand{\bb}{{\bi{b}}} 
\newcommand{\bze}{{\bzeta}} 
\newcommand{\bw}{{\bi{w}}} 
\newcommand{\M}{{\tilde M}} 
\newenvironment{pf}{\trivlist \item[\hskip\labelsep{\bf Proof.}]\rm}%
{\hfill$\Box$\endtrivlist}
\newlength{\myhspace}
\newcommand{\mhs}{\hspace*{\myhspace}} 
\newtheorem{thm}{Theorem}[section] 
\newtheorem{prop}[thm]{Proposition} 
 
\newtheorem{cor}[thm]{Corollary}

\begin{document} 
\jl{1}
\title[Sustained resonance]
  {Sustained resonance: a binary system perturbed by gravitational 
radiation} 
\author{C  Chicone\dag, B  Mashhoon\ddag\  and D G Retzloff\S}
\address{\dag\ Department of Mathematics,
University of Missouri,\\ Columbia, MO 65211, USA} 
\address{\ddag\ Department of Physics and Astronomy, 
University of Missouri,\\ Columbia, MO 65211, USA} 
\address{\S\ Department of Chemical Engineering, 
University of Missouri,\\ Columbia, MO 65211, USA} 
\begin{abstract} 
The general phenomena associated with sustained resonance 
are studied in this 
paper in connection with relativistic binary pulsars. 
We represent such a system by two point masses in a Keplerian binary
system that evolves via gravitational radiation damping as well as an
external tidal perturbation. 
For further simplification, we assume that the external 
tidal perturbation is caused by a normally incident 
circularly polarized monochromatic gravitational wave.
In this case, the second-order partially averaged equations are studied
and a theorem of C. Robinson is employed to prove that for certain
values of the physical parameters 
resonance capture followed by sustained resonance is possible
in the averaged system.
We conjecture that sustained resonance 
can occur in the physical system
when the perturbing influences nearly balance each other. 
\end{abstract} 
\pacs{0430,0540,9510C,9780}
\maketitle
\section{Introduction} 
Imagine a stable Hamiltonian system that is 
subjected to an external periodic perturbation as well as a 
frictional force that continuously takes energy out of the system.  
If the driving force vanishes on the average, then in general the 
average behavior of the system is characterized by damped motion.  
Under favorable conditions, however, the system could get
captured into a prolonged resonance if the driving frequency is 
commensurate with an internal frequency of the system.  In this paper, 
we study the dynamics of this phenomenon in the context of a simple 
model of certain relativistic binary systems that are of current 
interest.

Relativistic binary pulsars are generally expected 
to occur in some astrophysical environment and the gravitational effect of 
the surroundings on the binary orbit would be tidal in the first order of 
approximation. The evolution of the binary orbit under the combined 
influence of gravitational radiation reaction and the external tidal 
perturbation can lead to resonance under appropriate conditions [1--6]. 
Once we average over the external periodic tidal perturbation, the binary 
pulsar is expected to lose energy to 
gravitational radiation damping so that the relative orbit would spiral 
inward unless there is resonance capture. This would come 
about if the external source deposits energy into the orbit over an 
extended period of time so as to compensate for the continuous loss of 
energy due to radiation reaction. This situation, which persists until the 
system leaves the resonance, can occur if the Keplerian period of the 
osculating ellipse becomes commensurate with the period of the external 
perturbation. In this way, the binary pulsar can be captured into 
resonance with its gravitational environment. Then, the standard data 
reduction schemes are expected to reveal that during resonance capture the 
semimajor axis of the binary is not monotonically decreasing due to the 
emission of gravitational radiation but is instead oscillating about a 
fixed resonance value since the external perturbation continuously deposits 
energy into the orbit during resonance so as to replenish on average the 
energy lost to gravitational waves. The system leaves the resonance when 
this delicate energy balance is significantly disrupted. Such resonances 
are well known in the solar system, where the damping effects have a 
different origin (see \cite{NR1} and the references cited therein). One 
expects that over time the same kind of phenomenon 
will be observed in a sufficiently large sample of relativistic binary 
pulsars \cite {NR2}. 

It should be mentioned that there are transient chaotic phenomena that are 
generally associated with resonance. The transient chaos \emph{per se}
may be 
difficult to detect observationally due to noise and other complications 
that exist in the data reduction process \cite{smk}. On the other hand, 
there is numerical evidence for a peculiar chaotic effect, namely {\it 
chaotic transition}, that appears to be associated with transient chaos and 
involves a sudden transformation in the rate of orbital collapse 
\cite{6,cmr5}. That is, the system makes a transition from one relative 
orbit to another that collapses much more rapidly. Presumably such a 
chaotic transition---if it indeed occurs in nature---should be detectable 
by the timing observations of relativistic binary pulsars. 
Though the unperturbed binary systems in our investigations always involve 
standard astronomical masses on Keplerian orbits, it is possible to extend 
the analysis to geodesic orbits around black holes. Such general 
relativistic systems involve stable Kepler-like orbits as well as unstable 
orbits. We expect that our analysis of transient chaos could be extended 
to the stable orbits, while the occurrence of chaos involving the unstable 
orbits perturbed by incident gravitational radiation has been demonstrated 
by Bombelli and Calzetta \cite{lbec}. These results could be significant 
for the study of chaotic accretion onto black holes and possible chaotic 
enhancement in the emission of gravitational waves associated with active 
black holes.

The purpose of the present work is to show that resonance capture can 
indeed occur on average for 
a simplified model of the physical situation that has been 
described here, since our previous work on sustained resonance 
has been mainly numerical [2,4,5]. 
Our mathematical model is briefly discussed in the next section. 
The second-order partially averaged system is given in 
section~\ref{sec:pave}. 
The geometry of resonance capture for 
the second-order partially averaged system is 
presented in section~\ref{sec:gorc}, 
and it is shown that resonance capture 
can indeed occur for this partially averaged model under certain conditions. 
The proof of a resonance capture mechanism for the full model equations 
remains an important unsolved problem; 
instead, we provide numerical evidence for 
sustained resonance in our model in section~\ref{sec:discuss}. 
\section{Newton--Jacobi equation with radiation damping} \label{sec:newjac} 
In a series of papers [1--6], we have studied the long-term evolution of a 
Keplerian binary system coupled to a gravitational environment. In fact, 
the influence of the environment has been replaced---for the sake of 
simplicity and definiteness---by the tidal perturbation of an external 
gravitational wave. More precisely, we have considered the {\it simplest} 
gravitational model for a binary system under radiative perturbations. In 
the lowest (quadrupole) approximation, the external wave exchanges energy 
and angular momentum with the binary system but not linear momentum. 
For the purposes of the present investigation, we limit our considerations 
to a normally incident plane monochromatic external wave that is circularly 
polarized. The amplitude of this external perturbation is $\epsilon$, $0< 
\epsilon \ll 1$; it indicates the deviation of the spacetime metric from 
the Minkowski metric as a consequence of the existence of the incident 
radiation field. 
In practice, $\epsilon$ could be extremely small; in fact, direct
laboratory detection of gravitational waves from astrophysical sources with
$\epsilon\approx 10^{-20}$ is the goal of current experimental efforts.
The relative motion of the binary is planar due to the 
transversality of the external wave and is given by 
\begin{equation} 
\frac{{\rmd}^{2}x^{i}}{{\rmd}t^{2}} + \frac{\kappa x^{i}}{\rho^{3}}+{\cal 
R}^{i}=-\epsilon {\cal K}_{ij}x^{j}, 
\end{equation} 
where ${\bx}(t)={\bx}_{1}-{\bx}_{2}$ determines the 
relative orbit, the indices $i,j \in \{1,2\}$,  
$\rho=|{\bx}|$, and 
$\kappa =G_{0}(M_{1}+ M_{2})$. Here the binary system consists of two point 
masses $ M_{1}$ and $ M_{2}$ with the center of mass of the system at rest 
at the origin of spatial coordinates, i.e. 
${\bx}_{1}(t)= M_{2} \; {\bx}(t)/(M_{1}+M_{2})$ and ${\bx}_{2}(t)= - 
M_{1}\:{\bx}(t)/(M_{1}+M_{2})$. The radiation reaction term is given to 
lowest order by 
\begin{equation} 
\fl{\cal R} =\frac{4G^{2}_{0} \: M_{1} M _{2}}{5c^{5}\rho ^{3}}\Big[ 
\Big(12\rho ^{2} \dot {\vartheta}^{2}-18\dot{\rho}^{2} 
-\frac{4\kappa}{\rho}\Big)\dot{\bx} 
-\frac{\dot{\rho}} {\rho} 
\Big(36 \rho^{2}\dot{\vartheta}^{2}-14 \dot{\rho}^{2}+\frac{4\kappa}{3\rho}\Big) 
{\bx}\Big], 
\end{equation} 
where $\rho (t)$ and $\vartheta (t)$ describe the relative motion 
in terms of polar coordinates in the orbital plane and an overdot indicates 
differentiation with respect to $t$. The influence of the external 
radiation on the relative motion is represented in equation (1) by a 
symmetric and traceless tidal matrix ${\cal K}$ evaluated at the position 
of the center of mass, 
\begin{eqnarray} 
{\cal K} = \alpha \Omega^{2} \left[ \matrix{ 
\cos{ \Omega t} & \pm \sin{ \Omega t} \cr 
\pm \sin{ \Omega t} & - \cos{ \Omega t}} \right], 
\end{eqnarray} 
where $\Omega$ is the frequency of the external wave and 
$\alpha$, which is of order unity, is its amplitude. The upper (lower) 
sign indicates right (left) circularly polarized radiation in the 
transverse-traceless gauge. 
A general comment is in order here regarding the physical realizability
of such a perturbation. We can, for instance, imagine another
binary system with a relative orbit that is \emph{circular} with
Keplerian frequency $\Omega/2$. Then, in the rest frame of this
binary's center of mass and neglecting any dissipation the system emits
gravitational radiation of frequency $\Omega$ and the waves that propagate
perpendicularly to the orbital plane of the circular binary system are 
circularly polarized; far from the circular binary, these waves
are nearly planar and of the form under consideration here.

In the absence of damping, we have shown that if $\epsilon$ is sufficiently 
small, the planar relative motion 
is  bounded for all time as a consequence of the 
Kolmogorov-Arnold-Moser (KAM) theorem~\cite{5,cmr4}. 
The physical reason for this 
confinement is that the external wave does not monotonically deposit energy 
into the binary orbit; in fact, the general situation is that energy flows 
back and forth between the wave and the binary so that on average no net 
transfer of energy takes place. When the orbital damping is taken into 
account, the general tendency of the binary system is to collapse; however, 
under certain circumstances the system could get captured into resonance. 
The necessary condition for resonance is the commensurability of the 
Keplerian frequency $\omega$ and the wave frequency $\Omega$; that is, 
relatively prime integers $m$ and $n$ must exist for which $m \omega = n 
\Omega$. Sustained resonance would 
actually come about if a delicate balance could be established between the 
radiative perturbations.

It is useful to express the equations of motion in dimensionless form as in 
our previous work [3--5]. To this end, we express all spatial intervals in 
units of $R_{0}>0$ and all temporal intervals in units of $T_{0}>0$, 
where the scale parameters $R_{0}$ and $T_{0}$ are 
connected via 
$\kappa T_{0}^{2}/R_{0}^{3}=1$. Specifically, we assume that at 
some ``initial'' time the lengthscale $R_{0}$ is the semimajor axis 
of the binary 
orbit and $2 \pi T_{0}$ is its period. In terms of polar 
coordinates $(\rho, \vartheta)$, the equations of motion then take the form 
\begin{eqnarray}\label{eqcmr4} 
\dot{\rho}  =  {\cal P}_{\rho}, \nonumber \\ 
\dot{\vartheta}  =  \frac{{\cal P}_{\vartheta}}{\rho^{2}}, \nonumber \\ 
\dot{\cal P}_{\rho}  =  - \frac{1}{\rho^{2}} + \frac{{\cal 
P}_{\vartheta}^{2}}{\rho^{3}} + 4 \delta \frac{{\cal P}_{\rho}}{\rho^{3}} 
\left( {\cal P}_{\rho}^{2} + 6 \;\frac{{\cal P}_{\vartheta}^{2}}{\rho^{2}} 
+ \frac{4}{3\rho} \right) 
   - \epsilon \alpha \Omega^{2} \rho \cos{ (2 \vartheta \mp \Omega t)},
\nonumber  \\[.05in] 
\dot{\cal P}_{\vartheta} =  2 \delta \frac{{\cal 
P}_{\vartheta}}{\rho^{3}} \left( 9 {\cal P}_{\rho}^{2} - 6 \; \frac{{\cal 
P}_{\vartheta}^{2}}{\rho^{2}} + \frac{2}{\rho} \right) + \epsilon \alpha 
\Omega^{2} \rho^{2} \sin{ (2 \vartheta \mp \Omega t)}. 
\end{eqnarray} 
Here $\delta$ is the dimensionless parameter that characterizes 
radiation damping and is given by 
\begin{equation} 
\delta = \frac{4 G_{0}^{2} \: M_{1} M_{2}}{5 c^{5} T_{0} R_{0}}. 
\end{equation} 
For physical systems, $\delta < 2^{1/2}/40$ \cite{6}. We find that 
$\delta \simeq 10^{-15}$ for the Hulse--Taylor relativistic binary pulsar 
${\rm PSR}\;{\rm B}1913+16$; this same value for $\delta$ is also 
approximately valid for the relativistic binary pulsar ${\rm PSR}\;{\rm 
B}1534+12$ \cite{ihs}.

To study the general characteristics of system~\eqref{eqcmr4}, it is useful 
to express this system in terms of Delaunay's elements. These are the 
natural action-angle variables for the Kepler problem. At each instant of 
time $t$, the state of relative motion $({\bx}, \dot{\bx})$, or 
equivalently 
$(\rho,\; \vartheta, \; {\cal P}_{\rho}, \; {\cal P}_{\vartheta})$, defines 
an osculating ellipse that is tangent to the perturbed motion at $t$. The 
Delaunay variables $(\ell,\;\hat{g},\;L,\;G)$ are closely related to the 
orbital elements of the osculating ellipse. Let $a$ and $e$ be the 
semimajor axis and the eccentricity of the osculating ellipse, 
respectively; then, ${\cal P}_\vartheta^2=a(1-e^2)$,
${\cal P}_{\rho} {\cal P}_{\vartheta}=e \sin{ 
\hat{v}}$, and ${\cal P}_{\vartheta}^{2}/\rho=1+e \cos{ \hat{v}}$, where 
$\hat{v}$ is the true anomaly of the osculating ellipse. Moreover, let 
$\hat{u}$ be the corresponding eccentric anomaly; then, the Delaunay 
elements are defined by \cite{3,4} 
\begin{equation} 
\ell:=\hat{u}-e \sin{\hat{u}}, \; \hat{g}:=\vartheta - \hat{v}, \; 
L:=a^{1/2}, \; G:={\cal P}_{\vartheta}=L(1-e^{2})^{1/2} \;. 
\end{equation} 
Only positive square roots are considered throughout this work. 
Thus we limit our considerations to orbits with positive orbital angular 
momentum $G$. 

The Delaunay equations of motion 
can be written as 
\begin{eqnarray} \label{eqcmr7a} 
\dot{\ell} & =  \frac{1}{L^{3}}+\epsilon \: \frac{\partial {\cal H}_{\rm 
ext}}{\partial L}+\epsilon \Delta {\cal R}_{\ell}, \nonumber \\ 
\dot{\hat{g}}&  =  \epsilon \: \frac{\partial {\cal H}_{\rm ext}}{\partial 
G}+\epsilon \Delta {\cal R}_{\hat{g}}, \nonumber \\ 
\dot{L} & =  - \epsilon \: \frac{\partial {\cal H}_{\rm ext}}{\partial 
\ell}+\epsilon \Delta {\cal R}_{L}, \nonumber \\ 
\dot{G} & =  - \epsilon \: \frac{\partial {\cal H}_{\rm ext}}{\partial 
\hat{g}} + \epsilon \Delta {\cal R}_{G}, 
\end{eqnarray} 
where $\Delta = \delta /\epsilon$, ${\cal R}_{D}$, for $D \in 
\{\ell,\;\hat{g},\;L,\;G\}$, are radiation damping 
terms\footnote[1]{A misprint in the $\ell$-component of
the radiation damping terms in equation~(18) of~\cite{6}
and likewise equation~(A2) of \cite{7} should be corrected:
In the term proportional to $\frac{1}{r^2}$ inside the square bracket, 
the factor 
$(1+70 L^2/3-29 G^2/2)$ must be replaced by
$(70 L^2/3-27 G^2/2)$.} 
\cite{6,7}, and 
\begin{eqnarray} 
\fl{\cal H}_{\rm ext} = \frac{1}{2} \alpha a^{2}\Omega^{2} \Big\{\frac{5}{2} 
e^{2} \cos{ (2\hat{g} \mp \Omega t)} \nonumber \\ 
 + \sum_{\nu=1}^{\infty} \frac{1}{\nu} 
[ K_{+}^{\nu} (e) \: \cos{ (2 \hat{g} \mp \Omega t} + \nu \ell)  
+ K_{-}^{\nu} (e) \cos{(2 \hat{g} \mp \Omega t - \nu 
\ell)} ] \Big\}. 
\end{eqnarray} 
Here 
\begin{eqnarray} 
K_{\pm}^{\nu} (e)  :=  \frac{1}{2} \nu (A_{\nu} \pm B_{\nu}),\nonumber \\ 
A_{\nu}:=\frac{4}{\nu^{2} e^{2}}[2 \nu e (1-e^{2})J_{\nu}^{\prime}(\nu 
e)-(2-e^{2})J_{\nu}(\nu e)], \nonumber \\ 
B_{\nu}:=-\frac{8}{\nu^{2}e^{2}}(1-e^{2})^{1/2}\:[eJ_{\nu}^{\prime}(\nu 
e)-\nu (1-e^{2})J_{\nu}(\nu e)], 
\end{eqnarray} 
where $J_{\nu}(x)$ is the Bessel function of the first kind of 
order $\nu$ and $J_{\nu}^{\prime}(x)={\rmd}J_{\nu}(x)/{\rmd}x$. Finally, 
we note that $\Delta$ will be viewed as a fixed constant; that is, 
we will only consider perturbations in the $(\epsilon,\delta)$ parameter 
space along lines with equations of the form 
$\delta=\Delta\epsilon$. 

Let us note that in 
system~\eqref{eqcmr7a} there is only one ``fast'' angle, namely the mean 
anomaly $\ell$, and its frequency is the Kepler frequency $(\omega = 
L^{-3})$ of the osculating ellipse. Since ${\cal H}_{\rm ext}$ is 
explicitly time dependent, resonance occurs when $m\omega = n \Omega$, 
in which case $L$ becomes fixed and equal to $L_{\star}$. The dynamics of 
the system at the $(m:n)$ resonance is discussed in the next section. If 
the system is off resonance, however, the system~\eqref{eqcmr7a} can be 
averaged with the result that ${\cal H}_{\rm ext}$ averages out to zero and 
the binary system simply collapses due to gravitational radiation damping. 
\section{Partial averaging}
\label{sec:pave} 
The average value of ${\cal H}_{\rm ext}$ at the $(m:n)$ resonance is in 
general nonzero only for $n=1$; otherwise, the binary system simply 
collapses on average just as it does off resonance. Once the system is 
captured into a primary $(m:1)$ resonance, the resonance condition is not 
rigorously maintained. There are in fact deviations that have general 
characteristics. To study these, we let 
\begin{equation}\label{eqtransf} 
\ell =\frac{1}{L_{\star}^{3}}t+\phi , 
\qquad L=L_{\star}+\epsilon^{1/2}{\cal D}, 
\end{equation} 
where $\phi$ and ${\cal D}$ are new variables associated 
with resonance and $\epsilon^{1/2}$ is the corresponding small parameter. 
The average energy exchange due to the external perturbation is generally 
oscillatory 
\begin{equation}\label{eqcmr12} 
\langle{\cal H}_{\rm ext}\rangle = T_{c} \: \cos{ m \phi} + T_{s} \: \sin{ 
m \phi}, 
\end{equation} 
whereas the damping is unidirectional; in fact, we expect on  
the average slow oscillatory behavior about the resonance manifold.  
This motion can be described in terms of a general 
damped pendulum-like equation in $\phi$ with torque exerted by the 
radiation damping.

\begin{figure}[h!]
\centerline{\epsfbox{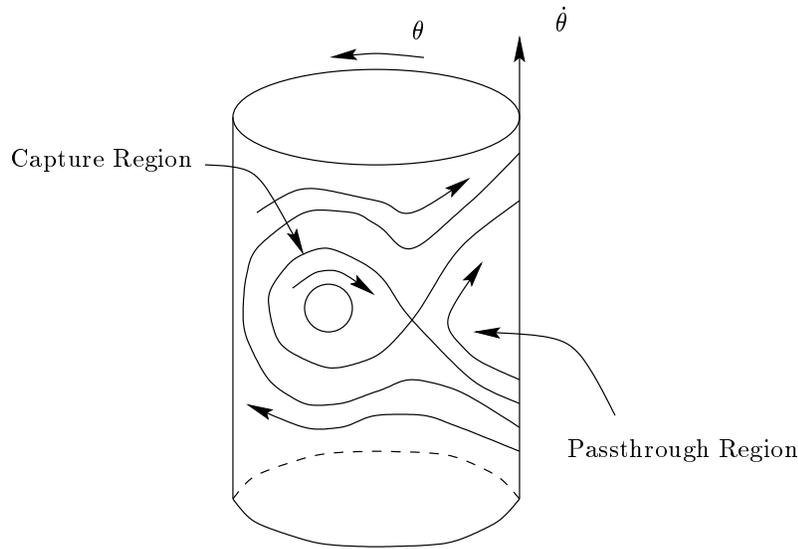}}
\caption{A schematic phase diagram on the phase cylinder for
the simple mathematical pendulum with 
constant torque $\ddot{\theta} - \lambda \sin{\theta} = \tau$
for the case $\lambda>\tau$.\label{Fig1}}
\end{figure}

The equations of motion averaged around the $(m:1)$ resonance and expressed 
to second order in $\epsilon^{1/2}$ have been given in \cite{7} for the 
planar Kepler system under radiative perturbations and in \cite{cmr5} for 
the corresponding three-dimensional case. Restricting the results of 
\cite{7} to the circularly-polarized incident wave under consideration here 
(see equations 18-19 of \cite{7}), the second-order partially averaged 
equations at the $(m:1)$ resonance are 
\begin{eqnarray}\label{eqcmr13} 
\dot{\phi}&=-\epsilon^{1/2}\frac{3}{L_{\star}^{4}}{\cal D}+\epsilon 
\Big(\frac{6}{L_{\star}^{5}}{\cal D}^{2}+ \frac{\partial T_{c}}{\partial L} 
\cos{ m \phi} +\frac{\partial T_{s}}{\partial L} \sin{ m \phi} \Big), 
\nonumber \\ 
\dot{\hat{g}}&=\epsilon \Big(\frac{\partial T_{c}}{\partial G} \cos{ m 
\phi} + \frac{\partial T_{s}}{\partial G}\sin{ m \phi} \Big), \nonumber \\ 
\dot{\cal D}&=-\epsilon^{1/2} \Big[ - mT_{c} \sin{ m \phi} + m T_{s} \cos{ 
m \phi} + \frac{\Delta}{G^{7}}(8+\frac{73}{3} e^{2}+\frac{37}{12}e^{4}) 
\Big] \nonumber \\ 
&\quad-\epsilon {\cal D} \Big[ -m\frac{\partial T_{c}}{\partial L} \sin{ m 
\phi} + m \frac{\partial T_{s}}{\partial L} \cos{ m \phi} + 
\frac{\Delta}{3L_{\star}^{3}G^{5}}(146+37e^{2}) \Big], \nonumber \\ 
\dot{G} & =  -\epsilon \Big[ \frac{\partial T_{c}}{\partial \hat{g}} \cos{ 
m \phi} + \frac{\partial T_{s}}{\partial \hat{g}} \sin{ m \phi} + 
\frac{\Delta}{L_{\star}^{3}G^{4}}(8+7e^{2}) \Big]. 
\end{eqnarray} 
Here $T_{c}=-\hat{f} \cos{2\hat{g}}$, $T_{s}=\pm \hat{f} 
\sin{2\hat{g}}$, and 
\begin{equation}\label{fhateq} 
\hat{f}:=\hat{f}(L,G)=-\frac{1}{2}\alpha m L_{\star}^{-6}L^{4}K_{\pm}^{m}(e), 
\end{equation} 
where $e=(1-G^2/L^2)^{1/2}$. 
The general behavior of the functions $K_{\pm}^{m}(e)$ has been 
discussed in \cite{cmr6} 
(see figure~\ref{Fig1} and appendix A of \cite{cmr6}).  
The functions $K_{-}^{m}(e)$ and $K_{+}^{1}(e)$ are both negative for $0<e<1$. 
In general, the derivative of the function $e\mapsto K_{-}^{m}(e)$ 
is negative for $0 < e < 1$ (see p. 107 of \cite{5}).   
Moreover, for $m>1$, 
$K_{+}^{m}(e)$ is positive for $0<e<\hat{e}_{m}$, zero for 
$e=\hat{e}_{m}$, and negative 
for $\hat{e}_{m}<e<1$; in fact, 
$\hat{e}_{2}\simeq 0.76$, $\hat{e}_{3}\simeq 0.85$, etc.

It is interesting to note that for incident radiation 
that is circularly polarized, 
equation~\eqref{eqcmr12} implies that $\langle{\cal H}_{\rm 
ext}\rangle=-\hat{f} \cos{(m \phi \pm 2\hat{g})}$. Introducing a new 
variable $\theta:=m\phi \pm 2\hat{g}$ and a new slow temporal parameter 
$\hat{t}:=-\mu t/b$, where $\mu:=\epsilon^{1/2}$ and 
$b:=L_{\star}^{4}/(3m)$, system~\eqref{eqcmr13} reduces to the form 
\begin{eqnarray}\label{decoup} 
\fl\dot{\theta}={\cal D}-\mu \Big[\frac{2}{L_{\star}}{\cal 
D}^{2}-b(m\hat{f}_{L} \pm 2\hat{f}_{G}) \cos{\theta} \Big], \nonumber \\ 
\fl\dot{\cal D}  =  mb\hat{f} \sin{\theta} + \frac{\Delta b}{G^{7}} 
(8+\frac{73}{3}e^{2}+\frac{37}{12}e^{4}) 
 + \mu b {\cal D} \Big[m\hat{f}_{L} \sin{\theta} 
+\frac{\Delta}{3L_{\star}^{3}G^{5}}(146+37e^{2}) \Big] , \nonumber \\ 
\fl\dot{G}  = \mu b \Big[ \pm 2\hat{f} \sin{ \theta} + 
\frac{\Delta}{L_{\star}^{3} G^{4}} (8+7e^{2}) \Big], 
\end{eqnarray} 
where the {\it overdot now refers to the new time} $\hat{t}$ and 
all relevant quantities are evaluated at 
$L=L_{\star}$. For example, 
$\left.\hat{f}_{L}=(\partial \hat{f}/\partial L)\right| _{L=L_{\star}}$, 
etc. 

It follows from equation~\eqref{decoup} that the first-order 
averaged equations reduce to $G = G_{\star}$, where $G_{\star}$ 
is a constant, and ${\cal D} = \dot{\theta}$, where $\theta(\hat{t\,})$ 
is a solution of 
\begin{equation} \label{pen16}
\ddot{\theta} - \lambda \sin{\theta} = \tau.
\end{equation}
Here $\lambda$ and $\tau$ are constants given by
\begin{eqnarray} \label{const17}
\lambda  = -\frac{1}{6} \alpha m L_{\star}^2 K^m_{\pm}(e_{\star}), \\
\label{const17a}
\tau  =  \frac{\Delta b}{G_{\star}^7}\big( 8+\frac{73}{3}e_{\star}^2 +
\frac{37}{12} e_{\star}^4 \big),
\end{eqnarray}
where $e_{\star} = (1-G_{\star}^2/L_{\star}^2)^{1/2}$.  
Equation~\eqref{pen16} describes a mathematical pendulum 
with a constant torque $\tau > 0$; the phase portrait of this 
dynamical system 
is illustrated in figure~\ref{Fig1} for $\lambda > \tau$.  
In this case, there is a homoclinic 
orbit which encloses a region corresponding to orbits 
that are captured into resonance.  
The region outside the homoclinic loop, with the stable and 
unstable manifolds removed, consists of orbits that pass through the
resonance.   The presence of the perturbation terms in the 
system~\eqref{decoup} implies that the pendulum is damped (or antidamped) 
in the second-order averaged equations.  That is, averaging the dynamical 
system to second order results in a pendulum equation of the 
form~\eqref{pen16} with slowly drifting $\lambda$ and $\tau$ as well as an 
additional term proportional to $\mu \dot{\theta}$.  The motion is 
geometrically described in the next section.

There are several important phenomena associated with resonance; for instance, chaos and resonance capture. 
Chaotic motion is expected to occur near a resonance under a Hamiltonian 
perturbation. For non-Hamiltonian perturbations, while chaotic motions 
are certainly possible, resonance capture is often the 
dominant phenomenon. At first order, the partially averaged 
equations at a resonance are generally pendulum-like equations with 
torque. This nonlinear Hamiltonian system can exhibit 
chaos and resonance capture under perturbation. The relationship 
between the chaos in the original dynamical system and possible chaotic 
effects in the averaged system near $(m:1)$ resonance is rather subtle and 
beyond the scope of this work. Here, we will consider instead the 
possibility of resonance capture. Indeed, we will give a geometric 
description of resonance capture in system~\eqref{decoup} and a rigorous 
proof that resonance capture does in fact occur.

\section{The geometry of resonance capture and a theorem of Robinson} 
\label{sec:gorc} 
We start from our model equations in Delaunay variables~\eqref{eqcmr7a} 
and introduce the transformation~\eqref{eqtransf} about a primary 
resonance; then, the resulting system can be transformed into the 
system~\eqref{decoup} plus terms of higher order by means of an 
averaging transformation.  These higher-order terms may be nonnegligible 
and dealing with this problem is beyond the scope of our work;
hence, we consider here resonance capture only in the second-order
partially averaged equations and not in the full model.    
System~\eqref{decoup} is typical of the following class of 
differential equations 
\begin{eqnarray}\label{robsys} 
\dot \by=F_0(\by,\bz)+\mu F_1(\by,\bz)+\Or(\mu^2),\nonumber\\ 
\dot \bz=\mu Z_1(\by,\bz)+\Or(\mu^2), 
\end{eqnarray} 
where $\by\in\bbR^2$ and $\bz\in \bbR^N$. In particular, 
note that $\bz$ is changing on a slow time scale relative to $\by$. For 
system~\eqref{decoup}, the pair $(\theta,\calD)$ plays the role 
of $\by$, and $G$, in our case a scalar variable, plays the role of $\bz$.

For the geometry of resonance capture, let us make a basic hyperbolicity 
assumption 
about system~\eqref{robsys}, namely, let us assume that 
for each fixed $\bze$ in some bounded region $\calZ$ of $\bbR^N$ the 
planar differential equation 
\begin{equation}\label{rsup1}
\dot \by= F_0(\by,\bze)
\end{equation} 
has a hyperbolic saddle point $P(\bze)$ with a corresponding homoclinic 
loop given by a solution $t\mapsto \by(t,\bze)$. Then, the set 
\[ M_0:=\{(\by,\bz): \by=P(\bz),\,\bz\in \calZ\} \] 
is a normally hyperbolic invariant manifold (consisting entirely of 
rest points) for the unperturbed system~\eqref{robsys}, i.e.  

\begin{figure}
\centerline{\epsfbox{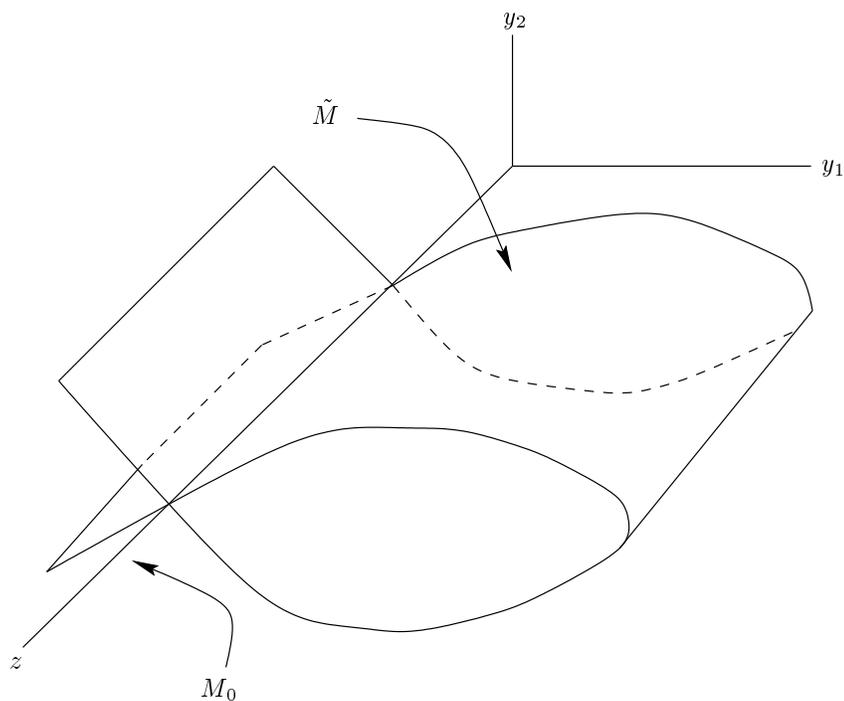}}
\caption{A homoclinic manifold for system~\eqref{rsup}.
\label{Fig2}}
\end{figure}
\begin{figure}
\centerline{\epsfbox{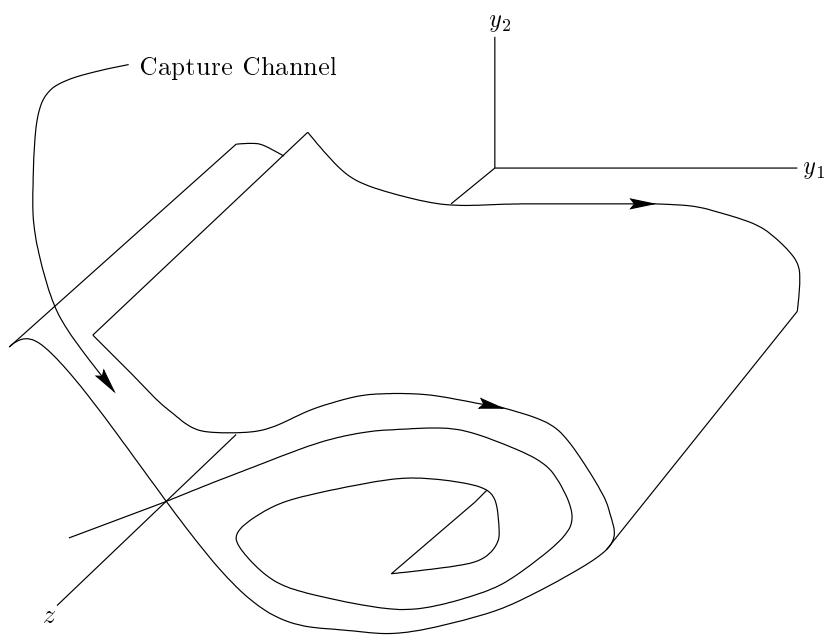}}
\caption{ 
The geometric configuration 
of the perturbed 
homoclinic manifold in figure~\ref{Fig1} 
corresponding to resonance capture. 
\label{Fig3}} 
\end{figure}
\begin{equation} \label{rsup} 
\dot \by= F_0(\by,\bz), \qquad \dot \bz=0. 
\end{equation} 
A rough description of this manifold is that nearby orbits are attracted or 
repelled from the vicinity of $M_0$ exponentially fast and the 
rates of expansion and contraction dominate the corresponding 
rates for orbits on $M_0$. Of course, in our case where the manifold 
consists of rest points, the second condition is satisfied automatically. 
The normally hyperbolic manifold $M_0$ forms the ``spine'' of the 
submanifold $\M$ in $\bbR^2\times\bbR^N$ 
consisting of the manifold $M_0$---given as a graph over an 
open subset of $\bbR^N$---and all the 
corresponding homoclinic loops as illustrated in figure~\ref{Fig2}. 
The manifold $\M$ ``bounds'' 
a region in the phase space for system~\eqref{rsup}, 
and this region corresponds to the initial conditions for orbits 
that are captured into resonance relative to system~\eqref{rsup}. 
When the system is subjected to perturbation, $\M$ 
is expected to ``split'', for example, as depicted in figure~\ref{Fig3}. 
In fact, by a celebrated theorem of 
N. Fenichel~\cite{fen}, the normally hyperbolic 
manifold $M_0$ persists. If the perturbation parameter $\mu$ is 
sufficiently small, then the perturbed system has an ``invariant'' 
normally hyperbolic manifold $M_\mu$ that is again given as a graph 
over a region in $\bbR^N$; in particular, the manifold $M_\mu$ 
has stable and unstable manifolds. Under appropriate conditions that 
will be specified below, the stable and unstable 
manifolds of $M_\mu$, the manifolds resulting from the 
split homoclinic manifold $\M$, 
form a channel in the phase space where 
orbits of the phase flow enter the vicinity of the resonance 
(cf. figure~\ref{Fig3}). This 
is the geometric description of resonance capture for a system 
such as~\eqref{robsys}.

Let us note that because 
the unperturbed ``slow manifold'' $M_0$ consists entirely of rest points, 
it is an invariant set for the unperturbed flow. However, 
the perturbed manifold $M_\mu$ generally has no fixed points; rather, 
the flow on $M_\mu$ is characterized by a slow drift in the $\bz$ variable. 
Thus, $M_\mu$ is only invariant in the sense that orbits with initial 
conditions on this manifold stay on it until they exit at one of 
its boundary points. 
Of course, as an effect of the slow drift, an orbit can, 
after a (perhaps long) sojourn near the resonance, enter a new dynamical regime. 
The sojourn time can be predicted from an analysis of the flow 
on the slow manifold. Thus, our geometric picture gives a rather 
complete description of the phenomena associated with resonance capture.

To prove that the geometric description of resonance capture obtains 
for system~\eqref{robsys}, we must detect splitting of $\M$ and determine 
the positions of the stable and unstable manifolds of $M_\mu$. 
This is accomplished by an adaptation of 
Melnikov's method that was first carried out by Robinson~\cite{cr}. 
To state his result, let us choose a portion of a plane $\calL$ in 
$\bbR^2\times\bbR^N$ that intersects each homoclinic loop in $\M$ 
perpendicular to the unperturbed vector field on $\M$. 
The choice of $\calL$ gives a smooth choice of initial conditions for 
the unperturbed orbits that correspond to the homoclinic loops. 
Indeed, for each $\bze$, consider the corresponding saddle point 
$P(\bze)$ and the unperturbed solution 
$t\mapsto \by_0(t,\bze)$ such that $\by_0(0,\bze)$ corresponds to the 
intersection point of $\calL$ with the homoclinic loop at $P(\bze)$. 
Let $d(\mu,\bze)$ denote the distance between the intersections 
of the stable and unstable manifolds of $M_\mu$ as measured along 
$\calL$ in the intersecting 
plane parallel to the $\by$ coordinate plane and passing 
through the point $(P(\bze),\bze)$. 
Also, for two vectors ${\ba},{\bb}\in\bbR^2$, 
let ${\ba}\wedge {\bb}:=a_1b_2-a_2b_1$. 
The following theorem is proved in~\cite{cr}: 

\begin{thm}\label{robth} 
If system~\eqref{robsys} satisfies the basic hyperbolicity assumption 
and the vector field given by $\by\mapsto F_0(\by,\bze)$ is 
divergence-free for each fixed $\bze\in \calZ$, 
then the splitting distance is given 
by $d(\mu,\bze)=\mu d_1(\bze)+\Or(\mu^2)$, where the leading-order 
coefficient is given by the integral 
\begin{equation}\label{robint} 
d_1(\bze)=\int_{-\infty}^\infty 
\Big( 
F_1+\frac{\partial F_0}{\partial\bz}\frac{\partial \bz}{\partial\mu} 
\Big) 
\wedge F_0\, \rmd t, 
\end{equation} 
all functions in the integrand are evaluated 
at $(\by_0(t,\bze),\bze)$, and the function 
\[t\mapsto \frac{\partial \bz}{\partial\mu}(\by_0(t,\bze),\bze)\] 
satisfies the variational initial value problem 
\[ \dot Y=Z_1(\by_0(t,\bze),\bze),\qquad Y(0)=0.\] 
\end{thm} 
\noindent The following immediate corollary is also stated in~\cite{cr}: 
\begin{cor} 
If $d_1(\bze)>0$ for all $\bze\in \calZ$, then there is resonance 
capture. Moreover, the only way a captured solution can leave the vicinity 
of the resonance is to have its second component, $\bz$, leave the region 
where $d_1(\bze)>0$, a process that occurs on a timescale of
order $1/\mu$. 
\end{cor}
\noindent The existence of 
resonance capture means that a capture channel opens 
up as in figure~\ref{Fig3} and therefore there is sustained 
resonance.

Let us now 
formulate and prove a version of 
another result of Robinson~\cite{cr} that 
we use to apply theorem~\ref{robth} to the pendulum-type systems 
that arise by partial averaging at resonance. 
Recall that system~\eqref{decoup} is derived from the averaged
physical model 
in part by a reversal of time. Thus, it is convenient to have a result 
that implies ``release from resonance'' for system~\eqref{decoup} so 
that it will imply capture into resonance for the original averaged model. 
Of course, the phrase ``release from resonance'' means 
that the stable and unstable manifolds of the normally hyperbolic 
manifold $M_0$ split in the direction that is opposite to that 
required for capture; in effect, orbits leave the vicinity 
of the resonance in a channel bounded by the unstable manifold.
 
\begin{prop}\label{rt} 
Consider the system 
\begin{eqnarray} 
\dot{\varphi}  = v +\mu \: p(\varphi ,v,\bw),\nonumber\\ 
\dot{v}  =  f(\bw) \sin{\varphi} +g(\bw) 
+\mu \: q(\varphi ,v,\bw),\nonumber \\ 
\dot{\bw}  =\mu \: r(\varphi ,v,\bw), \label{eq32} 
\end{eqnarray} 
where 
$(\varphi ,v,\bw) \in \bbR \times \bbR\times \bbR^N$, $\mu \geq 0$, and all 
indicated functions are smooth. If there is a point $\bw\in {\bbR}^N$ such 
that 
\begin{equation}\label{eq33} 
f(\bw)>0,\quad g(\bw) >0,\quad \frac{g(\bw)}{f(\bw)} <1, 
\end{equation} 
then the unperturbed system 
\[\dot{\varphi} =v, \quad \dot{v} =f(\bw)\sin{\varphi} +g(\bw)\] has a 
homoclinic loop that encloses a region $R(\bw)$ in the $(\varphi,v)$-plane. 
Suppose that, 
in addition, $t\mapsto (\varphi (t,\bw),v(t,\bw))$ is a solution 
on the homoclinic loop with 
\begin{eqnarray} 
f'(\bw)r(\varphi(t,\bw),v(t,\bw),\bw)\leq 0,\label{eq34}\\ 
\Big(\frac{g}{f}\Big)'(\bw)\; 
r(\varphi (t,\bw),v(t,\bw),\bw)\geq 0,\label{eq35} 
\end{eqnarray} 
where a prime denotes differentiation with respect to $\bw$, and 
for $(\varphi,v)\in R(\bw)$, 
\begin{equation}\label{eq36} 
p_\varphi(\varphi ,v,\bw)+q_v(\varphi ,v,\bw)\geq 0. 
\end{equation} 
If at least one of the inequalities~\eqref{eq34}, \eqref{eq35}, 
or~\eqref{eq36} is strict, then Robinson's integral $I_{\rm R}=d_1$ 
in~\eqref{robint} 
is negative at $\bw$. 
In particular, the perturbed configuration of stable and unstable manifolds 
corresponds to ``release from resonance''. 
\end{prop} 
\begin{pf} The first statement of the proposition, about the 
existence of homoclinic loops, is easily proved by elementary 
phase-plane analysis.
Let $s$ denote a new time-like variable given by 
\[\rmd s=(f(\bw))^{1/2}\: \rmd t,\] 
such that $s=0$ at $t=0$ and let 
$V:=v /(f(\bw))^{1/2}$. 
In these new variables, 
system~\eqref{eq32} has the 
form 
\begin{eqnarray*} 
\frac{\rmd\varphi}{\rmd s}  &=  V+\mu\frac{p(\varphi,(f(\bw))^{1/2} 
V,\bw)}{(f(\bw))^{1/2}},\\ 
\frac{\rmd V}{\rmd s} & = \sin{\varphi} 
+\frac{g(\bw)}{f(\bw)}+\mu 
\Big(\frac{q(\varphi,(f(\bw))^{1/2}V,\bw)}{f(\bw)}\\
&\quad -\frac{1}{2} 
(f(\bw))^{-\frac{3}{2}}f'(\bw)Vr(\varphi,(f(\bw))^{1/2} V,\bw)\Big),\\ 
\frac{\rmd\bw}{\rmd s} 
&=\mu \frac{r(\varphi,(f(\bw))^{1/2} V,\bw)}{(f(\bw))^{1/2}}. 
\end{eqnarray*} 
Robinson's integral $I_{\rm R}$ is then given by 
\[
\fl I_{\rm R} = \int_{-\infty}^\infty 
\Bigg\{ \Bigg( 
\begin{array}{c} 
pf^{-1/2} \\ 
qf^{-1} - \frac{1}{2}f^{-3/2}f'Vr 
\end{array} 
\Bigg) 
+\Bigg( 
\begin{array}{c} 
0 \\ 
\Big(g/f\Big)' 
\end{array} 
\Bigg) 
\bw_\mu \Bigg\} \wedge 
\Bigg( 
\begin{array}{c} 
V\\ 
\sin{\varphi}+g/f 
\end{array} 
\Bigg)\, \rmd s, 
\]
where $\bw_\mu:=\partial\bw/\partial \mu$.
We find that $I_{\rm R} = I_1+I_2+I_3 $, where 
\begin{eqnarray*} 
I_1:=\int_{-\infty}^\infty 
[pf^{-1/2} \left( \sin{\varphi} +g/f\right)-Vq/f ]\,\rmd s, \\ 
I_2:=\frac{1}{2}\int^\infty_{-\infty} f^{-\frac{3}{2}}V^2f'r\,\rmd s, \\ 
I_3:=-\int_{-\infty}^\infty V\big(g/f\big)'\bw_\mu\,\rmd s. 
\end{eqnarray*} 
Also, let us note that $\bw$ is constant on the 
separatrix loop and 
let us take the solution of system~\eqref{eq32} such that
the partial derivative $\bw_\mu$ 
vanishes at $s=0$ as required in the statement of theorem~\ref{robth}.
The first integral $I_1$ can be rewritten as follows: 
\begin{eqnarray*} 
I_1&=\int^\infty_{-\infty} 
\Big[p f^{-1/2}\big(\sin{\varphi}+g/f\big) 
-v f^{-1/2}q/f\Big]f^{1/2}\,\rmd t\\ 
&= \frac{1}{f}\int^\infty_{-\infty} [p(f\sin{\varphi}+g)-vq]\,\rmd t. 
\end{eqnarray*} 
Under the hypothesis~\eqref{eq33}, the 
homoclinic loop $\Gamma$ in the 
$(\varphi ,v)$-plane is parametrized 
with negative orientation relative to the usual 
orientation of the plane. 
Also, let us denote by $R$ the bounded region with 
boundary $\Gamma$ in the $(\varphi ,v)$-plane. By Green's 
theorem and the inequality~\eqref{eq36}, we have that 
\[ 
\fl I_1=\frac{1}{f} \int^\infty_{-\infty} 
(p\dot{v} -q\dot{\varphi})\,\rmd t 
=-\frac{1}{f}\int_\Gamma p\,\rmd v-q\,\rmd\varphi 
=-\frac{1}{f} \int_R(p_\varphi +q_v)\,\rmd\varphi\, \rmd v \leq 0. 
\] 
\noindent It follows from hypothesis~\eqref{eq34} that 
$I_2\le 0$.
For the integral 
$I_3$, let 
us observe that if we take the 
parametrization of $\Gamma$ such that 
$V(0)=0$, then $sV(s)\geq 0$; that is, 
$V(s)$ has the same sign as the variable $s$ in the 
integration. Also, because 
$\bw$ is constant for the integration, 
$\mu=0$ for the functions in the integrand, and hypothesis~\eqref{eq35} 
holds, 
we have the inequality 
\[
\frac{\rmd}{\rmd s} \Big [ \Big( 
\frac{g}{f}\Big)'\bw_\mu \Big] =\Big(\frac{g}{f}\Big)' 
\frac{\rmd\bw_\mu}{\rmd s}\Big|_{\mu =0} 
=\Big(\frac{g}{f}\Big)' r f^{-1/2}\geq 0.
\] 
Thus, using the fact that $\bw_\mu$ vanishes at $s=0$, note that 
$(g/f)' \bw_\mu$ has the same sign as the variable $s$ for the 
integration. We conclude that 
\[V\Big( \frac{g}{f}\Big)'\bw_\mu \geq 0 \] 
over the range of the integration. Thus, $I_3\leq 0$. 

Finally, note that if  inequality~\eqref{eq36} 
is strict, then
$I_1<0$. Likewise, if inequality~\eqref{eq34} (respectively
inequality~\eqref{eq35}) is strict, then
$I_2<0$ (respectively $I_3<0$). Hence, $I_{\rm R}=I_1+I_2+I_3\le 0$
and if at least one of the inequalities~\eqref{eq34},
\eqref{eq35}, or~\eqref{eq36} is strict, then $I_{\rm R}<0$.
\end{pf}

To apply proposition~\ref{rt}, let us consider 
the second-order averaged equations for 
the normally incident left circularly polarized gravitational wave 
expressed in decoupled form; 
namely, system~\eqref{decoup} with the ``lower'' signs. 
Also, let us note that this system is a special case of the 
abstract system in proposition~\ref{rt} once we set 
$\theta=\varphi$, $\calD=v$, and $G=\bw$. 
Recall that if $L_{\star}>G>0$, then $0<e<1$ and 
$K^m_-(e)<0$. Hence, we have the desired inequality 
$\hat{f}(L_{\star},G)>0$ once we choose $\alpha > 0$.  In fact, 
we set $\alpha = 1$. 
We claim that the hypotheses of 
proposition~\ref{rt} are satisfied under the following two conditions: 
\begin{enumerate} 
\item[\quad 1.] There is a 
subinterval $\calG_1$ of the interval $(0,L_\star)$ such that 
\[
\frac{\Delta }{m\hat{f}G^7}\left( 
8+\frac{73}{3}e^2+\frac{37}{12}e^4\right) 
<1 
\]
whenever $G\in \calG_1$. 
\item[\quad 2.] 
There is an interval $\calG_2\subseteq \calG_1$ such that the derivative of 
the function 
\[
e\mapsto \frac{1}{(1-e^2)^{7/2}\hat{f}} 
\Big(8+\frac{73}{3}e^2 +\frac{37}{12}e^4\Big) 
\]
is negative on the image of the interval 
$\calG_2$ under the transformation 
\[e=(1-G^2/L^2_{\star})^{1/2}.\] 
\end{enumerate}
Condition~1 ensures that hypothesis~\eqref{eq33} is 
satisfied for each $G\in \calG_1$. 
Recall that 
\[ 
\hat{f}_G:=\frac{\partial \hat{f}}{\partial e} \frac{\partial e}{\partial 
G} = -\frac{1}{Ge}(1-e^2)\frac{\partial \hat{f}}{\partial e}, 
\] 
and since $\partial K^m_-(e)/\partial e < 0$ for $0 < e < 1$
we have that $\partial \hat{f}/\partial e >0$. 
Thus, $\hat{f}_G<0$ for $G\in \calG_1$. 
Turning to the function $r$ in the 
statement of proposition~\ref{rt}, note that 
\[
r:=-2b \hat{f}\sin{\theta} +\frac{\Delta b}{L^3_{\star}G^4} (8+7e^2). 
\]
By the first implication of proposition~\ref{rt}, 
it follows that if $G\in \calG_1$, then the 
system 
\[ 
\dot{\theta }=\calD,\quad 
\dot{\calD}=mb\hat{f} \sin{\theta} 
+\Delta \frac{b}{G^7}\Big( 8+\frac{73}{3}e^2+\frac{37}{12}e^4\Big) 
\]
has a homoclinic orbit where the 
corresponding saddle point has 
coordinates $(\theta ,\calD)=(\theta _{\star},0)$ for some $\theta_\star$ 
in the interval $(3\pi/2,\,2\pi)$ as long as $\theta \in [0,2\pi)$. 
At the boundary 
of the interval $\calG_1$, where 
\[
\frac{\Delta }{m\hat{f} G^7} 
\left(8+\frac{73}{3}e^2+\frac{37}{12}e^4\right) 
=1, 
\] 
the corresponding saddle point with coordinates $(\theta _{\star},0)$ 
and center with coordinates $(3\pi -\theta _{\star},0)$ 
coalesce at $(3\pi /2,0)$; that is, these rest points 
disappear in a saddle node bifurcation. 
Thus, there is a subinterval 
$\calG_2\subseteq \calG_1$ 
such that if $G\in \calG_2$, then the 
region $R(G)$, enclosed by the 
homoclinic orbit, contains only points 
whose first coordinates $\theta $ are in 
the interval $\pi <\theta <2\pi$. In 
particular, if $G\in \calG_2$, 
then $\sin{\theta} <0$ and therefore $r>0$. Hence, 
$r\hat{f}_G<0$ and 
hypothesis~\eqref{eq34} is satisfied.

To verify hypothesis~\eqref{eq35}, we must show that if 
$G\in \calG_2$, then the derivative of 
the function 
\[ 
G\mapsto \frac{\Delta}{m \hat{f}G^7} 
\Big(8+\frac{73}{3}e^2+\frac{37}{12}e^4\Big) 
\] 
is positive; however, this requirement follows immediately from 
condition 2.
For hypothesis~\eqref{eq36}, let us 
determine the sign of the divergence of the vector field $\calV$ given by 
\[\fl 
(\theta,\calD) \mapsto 
\Big[ 
-\Big(\frac{2\calD^2}{L_{\star}}+b(2\hat{f}_G-m\hat{f}_L)\cos{\theta}\Big),\: 
b\calD 
\Big(m\hat{f}_L\sin{\theta}+\frac{\Delta}{3L^3_{\star}G^5}(146+37e^2)\Big) 
\Big]. 
\] 
Indeed, using the inequalities 
$\sin{\theta} <0$ and $\hat{f}_G<0$, 
it follows that if $G\in \calG_2$, 
then the divergence of $\calV$ is the positive quantity 
\[ 
b\Big(2\hat{f}_G\sin{\theta}+\frac{\Delta}{3L^3_{\star}G^5}(146+37e^2)\Big). 
\] 
This completes the proof of our claim.

To show the desired phenomenon of release from resonance, we will verify 
conditions~1 and~2.
In order to determine the interval 
$\calG_2$, let us note that our (unperturbed) 
pendulum-type planar system has the form 
\begin{equation}\label{abpen} 
\dot{\theta}=\calD,\quad \dot{\calD}=\lambda \sin{\theta} + \tau, 
\end{equation} 
where $\lambda >0$, $\tau >0$, and $\tau/\lambda <1$. 
System~\eqref{abpen} is Hamiltonian with total energy 
\[H=\frac{1}{2}\calD^2 +\lambda \cos{\theta} -\tau \theta.\] 
Denote 
the coordinates of its hyperbolic saddle point by 
$(\theta _{\star},0)$, and recall that 
$3\pi /2<\theta _{\star}<2\pi$. Also, note 
that the homoclinic loop is given by the 
graphs of the functions 
\[
\calD=\pm 2^{1/2} 
(\lambda \cos{\theta_{\star}}-\tau \theta _{\star}-
(\lambda \cos{\theta} -\tau \theta ))^{1/2}.
\] 

\begin{prop}\label{p1} For system~\eqref{abpen}, 
there is a point $\hat{\theta}$ such that $\theta _{\star} 
-2\pi<\hat{\theta}<\theta_{\star}$ 
and 
\[ 
\lambda \cos{\hat{\theta}}-\tau \hat{\theta} =\lambda 
\cos{\theta_{\star}}-\tau \theta_{\star}. 
\] 
The point $(\hat{\theta }, 0)$ is not a rest point. 
Moreover, the equation 
\[ 
\lambda \cos{\theta} -\tau \theta = 
\lambda \cos{\theta_{\star}}-\tau \theta_{\star}
\]
has no solution for $\theta >\theta_{\star}$. That is, the 
homoclinic loop at $(\theta_\star,0)$ crosses the $\theta$-axis 
at $\theta=\hat{\theta}$ and encloses that portion of 
the $\theta$-axis given by 
$\hat{\theta }\leq \theta \leq \theta_{\star}$. Also, 
$\hat{\theta }>\pi$ if and only if 
$k:=\tau/\lambda$ is such that 
\[\frac{1+(1-k^2)^{1/2}}{k} +\sin^{-1}k<\pi.\]
\end{prop} 
\begin{pf} Define the function $h$ by 
\[
h(\theta)=\lambda \cos{\theta}-\tau \theta 
-(\lambda \cos{\theta_\star}-\tau \theta_\star) 
\]
and note that it has relative maxima at $\theta=\theta_\star$ and 
$\theta=\theta_\star-2\pi$. Also, $h(\theta_\star)=0$ and 
$h(\theta_\star-2\pi)=2\pi \tau>0$. It follows that 
there is a point $\hat{\theta}$ in the interval indicated in the statement 
of the proposition such that $h(\hat{\theta})=0$. 
Moreover, this point is not a rest point. Indeed, the only rest point 
in the open interval $(\theta_\star-2\pi,\theta_\star)$ is 
the point $3\pi/2-\theta_\star$, and it corresponds to a negative 
relative minimum of $h$. 
If $\theta>\theta_\star$, and $h(\theta)=0$, then there must 
be some relative maximum of $h$ that is nonnegative. However,
all the relative maxima that exceed $\theta_\star$ 
have the form $\theta_{\star}+2\pi N$, where $N$ is a positive integer, 
and $h$ takes negative values at all such points; therefore,  we reach a 
contradiction.
Clearly, $\hat{\theta }>\pi$ if and only if $h(\pi) > 0$ or 
\begin{equation}\label{linq} 
\tau (\theta _{\star} - \pi) - \lambda (1+\cos{\theta_{\star}})>0. 
\end{equation} 
Recall that $\sin{\theta_{\star}}=-\tau/\lambda$ and 
$\cos{\theta_{\star}}>0$. Thus, with 
$k:=\tau/\lambda$,
inequality~\eqref{linq} is equivalent to the inequality 

\[\frac{1+(1-k^2)^{1/2}}{k}+\sin^{-1}k< \pi,\] 
as required. 
\end{pf}

In summary, let us recall that  $f(e) = mb \hat{f}(L_{\star},G)$, where $e = 
(1-G^2/L^2_{\star})^{1/2}$ and $\hat{f}(L_{\star}, G) = -(m/2L_{\star}^2)
K^m_{\pm}(e)$ by equation~\eqref{fhateq}.   
For incident left circularly polarized 
radiation, we have that $f(e)>0$ and $f'(e)>0$; in fact,
\[K^m_-(e) = - \frac{(5m+2)m^{m-1}}{2^{m+1}(m+2)!} e^{m+2} +\Or(e^{m+4})\: ,\]
and for $0 < e \leq 0.5$ the first term in this Taylor 
expansion is accurate to about 10\%.   
Then, all the hypotheses of 
proposition~\ref{rt} are satisfied provided that 
there is some $e$ such that $0<e<1$ and 
\begin{eqnarray*} 
{\rm (i)}  \quad 
k(e) =\frac{\Delta}{mL^7_{\star}} \: 
\frac{1}{f(e)(1-e^2)^{7/2}}\: \Big( 
8+\frac{73}{3}e^2+\frac{37}{12} 
e^4\Big) <1,\\ 
{\rm (ii)}  \quad k'(e)\leq 0,\\ 
{\rm (iii)}  \quad 
\frac{1+(1-k^2(e))^{1/2}}{k(e)} +\sin^{-1}{(k(e))}<\pi. 
\end{eqnarray*}

\noindent One can show by inspection that condition (iii) is 
satisfied provided that $k_0 <k(e)\le1$, where $k_0 \approx 0.725$. 
We therefore choose $k(e) = 0.9$ in agreement with condition (i).  
Moreover, one can directly verify that $k^{'}(e) < 0$ if we assume 
$e = 0.5$ and $m = 2$.  It is interesting to note that the 
eccentricities of the relativistic binary pulsars PSR B1913+16 and 
PSR B1534+12 are $\approx 0.6$ and $\approx 0.3$, respectively.  
We choose $L_{\star} = (4/3)^{1/2}$, so that $\Omega \approx 1.3$ 
at the (2 : 1) resonance that we wish to consider.  
Finally, $\Delta$ is uniquely determined from $k(e) =0.9$ once we recall that 
$\alpha = 1$ and $K^2_{-}(0.5) \approx - 0.009$; 
hence, $\Delta \approx 5 \times 10^{-4}$.  
\section{Discussion} \label{sec:discuss}
We have shown that a set of parameters exists such that for incident 
left circularly polarized radiation the averaged equations permit 
sustained resonance. 
The essential requirement for sustained resonance in our approach
is that the Robinson integral be nonzero. Our objective has been
to \emph{prove} sustained resonance in the second-order averaged equations  
for our model. However, we expect that Robinson's integral is
generally nonzero without the restrictive assumptions of 
proposition~\ref{rt}. Indeed, it should be clear from the 
proof of proposition~\ref{rt} that $I_{\rm R}$ can be negative
without the requirement that each of the three integrals
$I_1$, $I_2$, and $I_3$ be negative. For this reason, we 
expect that resonance capture would generally be possible for either circular 
polarization state of the incident radiation.  Moreover, circular 
polarization is not essential for sustained resonance as demonstrated 
by our previous numerical work on resonance capture that did not 
involve circular polarization [2,4,5].  Indeed, the mathematical theory of 
sustained resonance developed in this paper owes 
much to the simplification---i.e.\  the 
reduction from four to three phase-space dimensions in the 
system~\eqref{decoup}---brought about by the assumption that the 
normally incident radiation has definite helicity.

\begin{figure}[hp!]
\centerline{\epsfbox{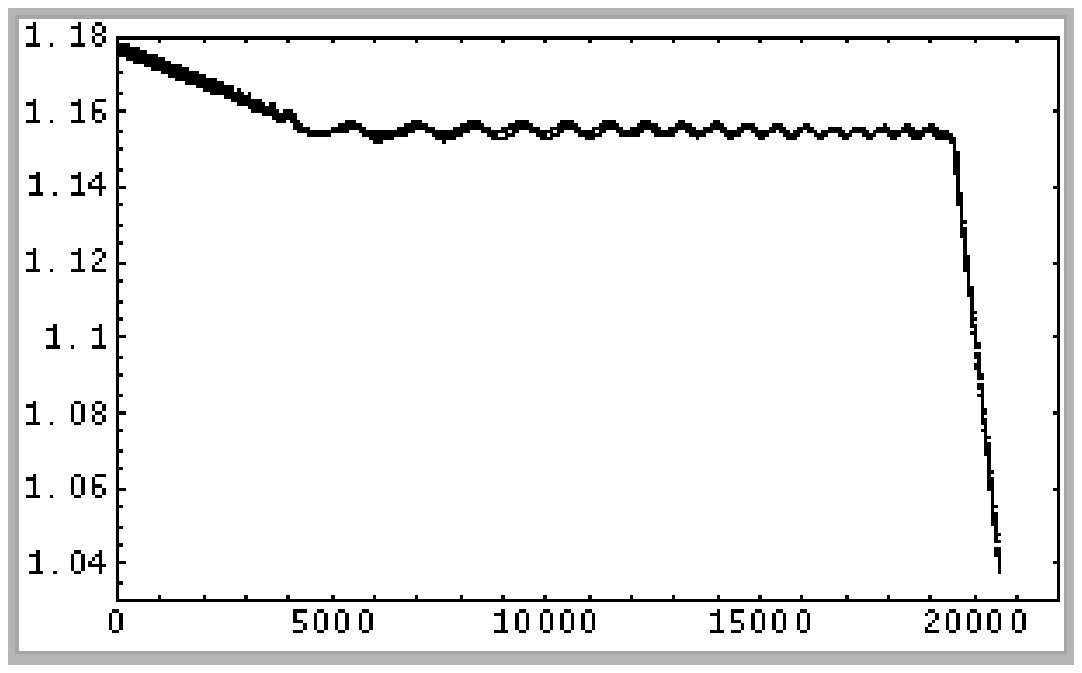}}
\centerline{\epsfbox{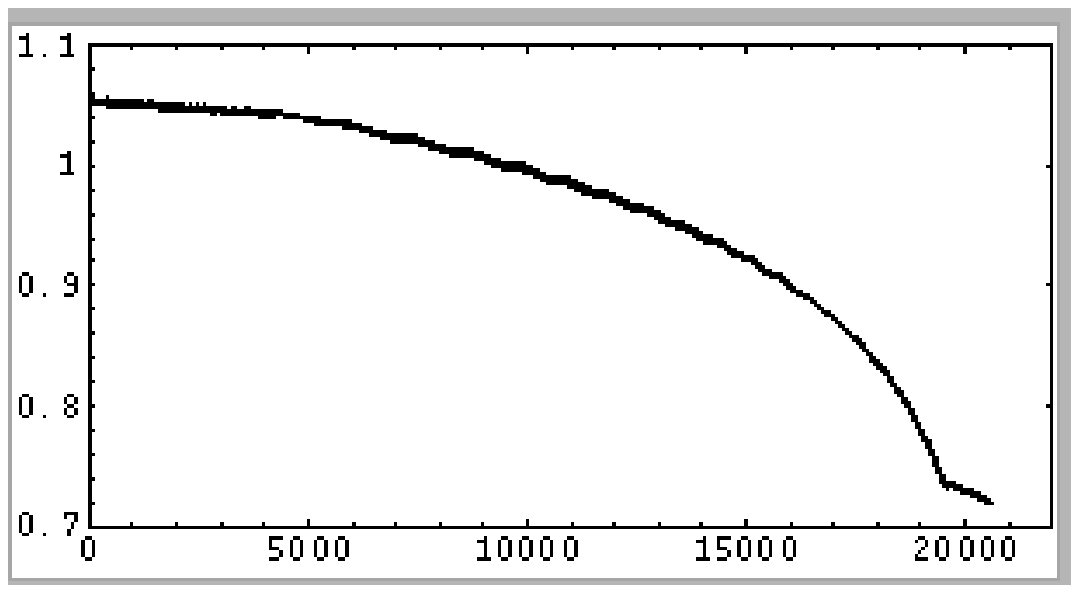}}
\centerline{\epsfbox{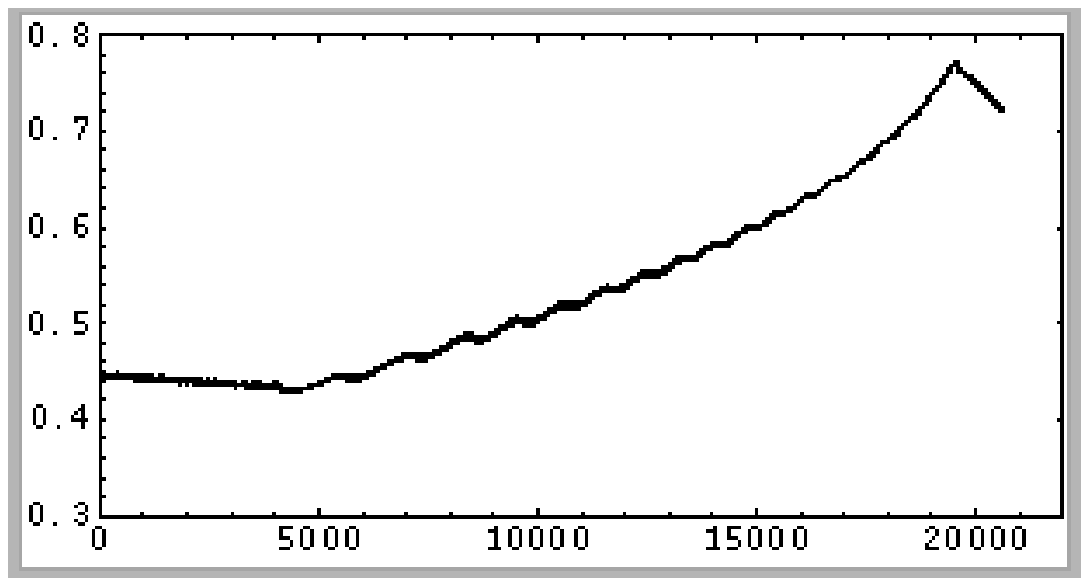}}
\caption{Sustained resonance for the relative orbit in a binary system 
with initial conditions $(\rho, \vartheta, \calP_{\rho}, \calP_{\vartheta})
 = (1,0,0.5,1)$ in (2:1) resonance with a normally incident 
left circularly polarized gravitational wave.  
In the graphs, produced by backward and forward integration 
as explained in the text, 
this initial point corresponds to $t\approx 9500$. 
Here $\alpha = 1$, $\epsilon = 10^{-3}$, $\delta = 5\times 10^{-7}$, and 
$\Omega = 2(3/4)^{3/2} \approx 1.299$.  The top, middle, and
bottom panels depict, respectively, $L$, $G$, and $e$ versus time.  
The rate of orbital decay after the resonance is generally different 
from that before the resonance, since during resonance capture the 
orbit exchanges angular momentum with the incident radiation 
such that the orbit has a different eccentricity when it 
exits the resonance.  The duration of the resonance 
is $\approx 15/\epsilon$ time units, in agreement with corollary 4.2.
\label{Fig4}}
\end{figure}
\begin{figure}[hp!]
\centerline{\epsfbox{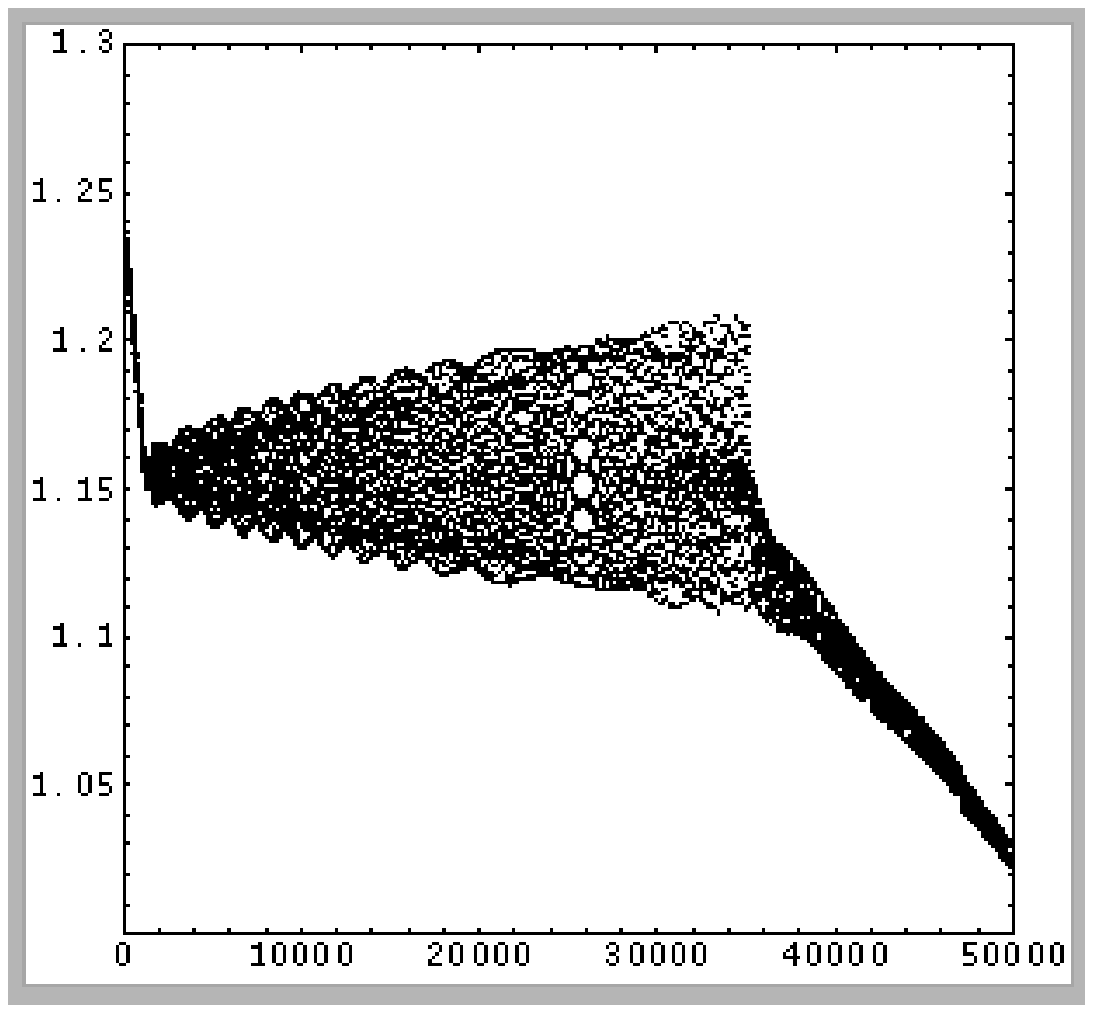}}
\centerline{\epsfbox{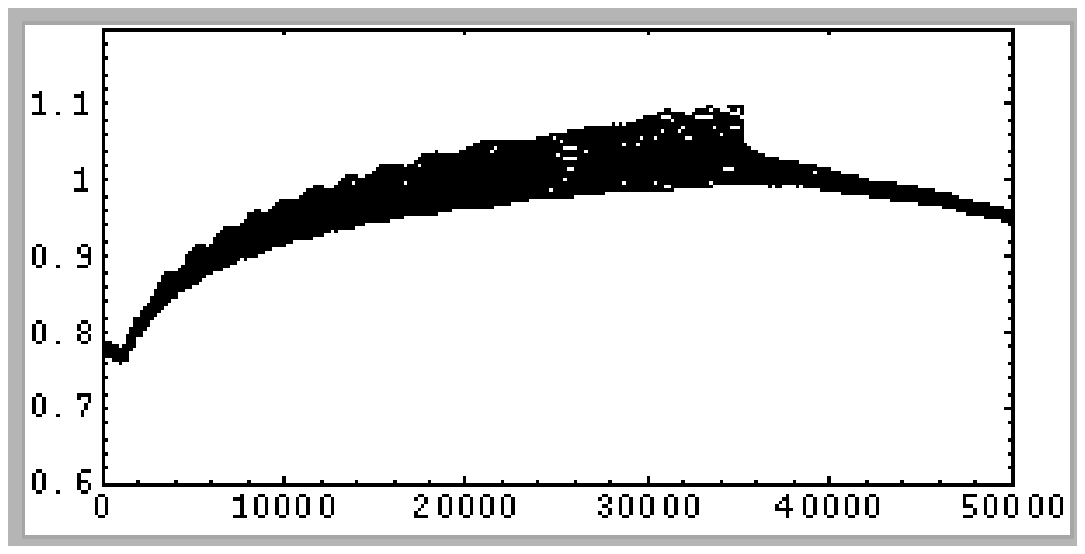}}
\centerline{\epsfbox{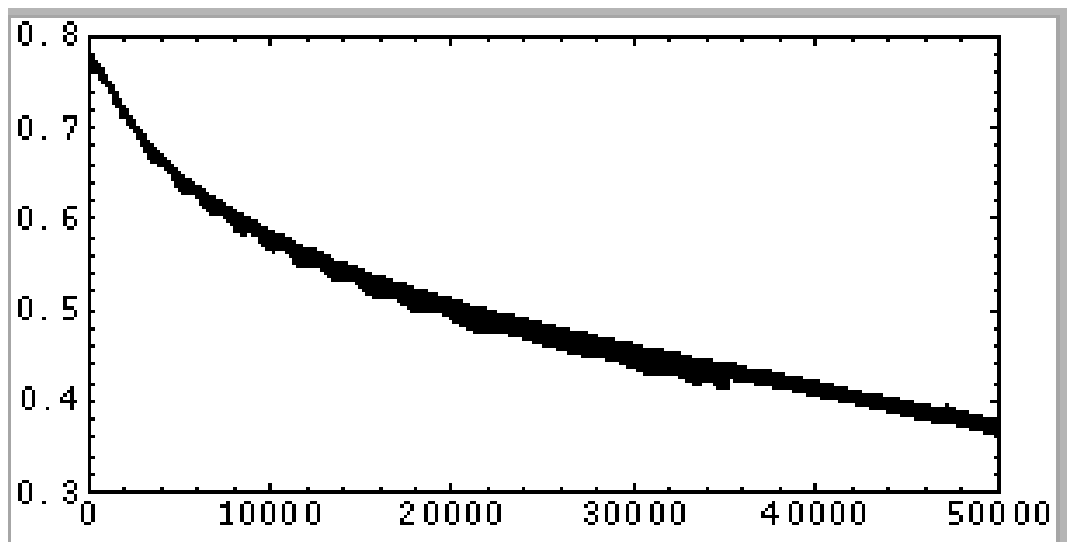}}
\caption{Same as in figure~\ref{Fig4} 
except that the incident radiation 
has  positive helicity. The duration of the resonance 
is $\approx 34/\epsilon$ time units; this is consistent  
with corollary~4.2.  The response of the system is more pronounced 
in this case, since the sense of the orbital angular momentum is the 
same as the helicity of the incident radiation. \label{Fig5}}
\end{figure}

The proof of the existence of sustained resonance involves the second-order  
partially averaged system.  We conjecture that the result can be extended 
to the original system.  To provide evidence for this conjecture, it 
is interesting to use the set of parameters obtained in the previous 
section as part of the initial conditions needed to integrate the 
original system numerically.  To this end, we integrate 
system~\eqref{eqcmr4} with initial conditions 
$(\rho, \vartheta, \calP_{\rho}, \calP_{\vartheta}) = (1, 0, 0.5, 1)$, 
$\Omega =2(3/4)^{3/2}$, $\epsilon = 10^{-3}$, and 
$\delta = \epsilon \Delta = 5\times 10^{-7}$.  Thus the initial orbit has 
$L = (4/3)^{1/2}$, $e = 0.5$, $G = 1$, and 
$\theta = 5\pi/3- {3}^{1/2}/2$, so that $\pi < \theta < 2\pi$ as required.  
The periastron for this initial osculating orbit, which has true
anomaly $\hat {v}=\pi/2$, occurs at
$\vartheta=-\pi/2$; hence $\hat{g}=-\pi/2$. In fact, this 
initial orbit has been used frequently
in our previous work [3--6]. 
We find that with these initial conditions the system is indeed in 
sustained (2 : 1) resonance for either circular polarization state
as demonstrated in figures~4 and~5.  
Figure~\ref{Fig4} can be produced as follows: System~\eqref{eqcmr4} is
integrated backward 
from the initial point for normally incident radiation that is 
left circularly polarized until the system leaves the resonance and
a final point of integration is stored; then, the graph is made by
integrating forward in time from this point.
It is interesting to note that the binary orbit collapses rather rapidly
following exit from  resonance when the orbit is highly eccentric. 
This is explained by the
fact that the rate of emission of gravitational radiation by the
binary orbit is proportional to $(1-e^2)^{-7/2}$; hence, as
$e\to 1$ the loss of orbital energy to radiation leads to
orbital collapse~\cite{6}.
Figure~\ref{Fig5} can be produced in a similar way and
illustrates the same 
situation for right circular polarization.  
The \emph{possibility} of resonance in this case can be seen
from the first-order averaged equations~\eqref{pen16}--\eqref{const17a};
that is, $K^2_+(0.5)\approx 0.923$ and
hence $|\lambda|>\tau$, so that the situation depicted in figure~1
holds except that the direction of some arrows should be reversed.
It is clear from figures~4 and~5 that evolutionary dynamics of the orbit
while trapped in resonance depends sensitively upon the helicity of 
the incident radiation. For example, the orbit loses angular momentum
while in resonance and becomes highly eccentric 
when the incident radiation has negative helicity as in figure~4, 
while the opposite situation holds in figure~5 for positive
helicity incident radiation; this circumstance may be naturally interpreted
in terms of the (negative or positive) angular momentum 
carried by the (negative or positive helicity) radiation and deposited
into the binary orbit.
The fact that the 
amplitude of oscillations around resonance is generally larger when the 
sense of the helicity of the radiation is the same as the rotation of the 
orbit is in agreement with our previous work on the Hill system.
In contrast to the Hill system,
the present case involves dissipation; nevertheless, the analysis
of the amplitude of oscillations around the resonance is very
similar to that carried out in section~6 of~\cite{cmr6} due
to the fact that $\tau<|\lambda|$.

Figures 4 and 5 also illustrate the complex structure that is usually
associated with a higher-order resonance~\cite{7}; furthermore,
the presence of chaos is strongly indicated in our numerical work.
Chaos is especially noteworthy in figure~5; therefore, the qualitative
significance of our numerical results should be emphasized. We have
performed numerical experiments with $\epsilon$ and $\delta$ twice
as large as those in figures~4 and~5. In this case, we find in the experiment
corresponding to figure~5 that the $(2:1)$ resonance appears to have
significant overlap with the $(3:1)$ resonance.
Let us note that we expect $\epsilon$ to be
many orders of magnitude smaller in
physical situations of interest; however, we have taken $\epsilon$ to be
$10^{-3}$ in figures~4 and~5 for the sake of simplicity.

Do all relativistic binary orbits {\it monotonically} decay by gravitational  
radiation damping?   There is evidence for such energy loss from the timing 
analysis of only two pulsar systems \cite{ihs}.  However, such a system 
could in principle fall into sustained resonance with an external 
source; during 
resonance lock, the orbit would not decay on the average.  The orbital 
decay would resume once the system is released from resonance.  
It remains to see---as more binary pulsars are discovered and their timing data 
accumulate---whether sustained resonance in fact 
occurs in relativistic binary systems.

\ack
Carmen Chicone was supported by the NSF grant 
DMS-9531811 and the University of Missouri Research Board.


\section*{References}
\begin{thebibliography}{99}
\bibitem{5} Chicone C, Mashhoon B and Retzloff D G 1996 
Gravitational ionization: periodic orbits of binary systems 
perturbed by gravitational radiation 
{\em Ann. Inst. H.~Poincar\'e, Phys. Th\'eor.} {\bf 64} 87--125
\bibitem{cmr4} Chicone C, Mashhoon B and Retzloff D G 1996 
On the ionization of a Keplerian binary system by periodic gravitational 
radiation {\em J. Math. Phys.} {\bf 37} 3997--4016; 1997 Addendum {\em 
ibid.} {\bf 38} 544
\bibitem{6} Chicone C, Mashhoon B and Retzloff D G 1997 
Gravitational ionization: a chaotic net in 
the Kepler system 
{\em Class. Quantum Grav.} {\bf 14} 699--723
\bibitem{7} Chicone C, Mashhoon B and Retzloff D G 1997 
Evolutionary dynamics while trapped in 
resonance: a Keplerian 
binary system perturbed by gravitational radiation 
{\em Class. Quantum Grav.} {\bf 14} 1831--1850

\bibitem{cmr5} Chicone C, Mashhoon B and Retzloff D G 1999 
Chaos in the Kepler system {\em Class. Quantum Grav.} {\bf 16} 507--527
\bibitem{cmr6} Chicone C, Mashhoon B and Retzloff D G 1999
Chaos in the Hill system {\em Helv. Phys. Acta},  in press
\bibitem{NR1} Henrard J 1985 Resonance sweeping in the solar system 
\emph{Stability of the Solar System and its Minor Natural and Artificial 
Bodies} ed V G Szebehely (Dordrecht: Reidel) pp 183--192
\newline \mhs 
Beaug\'e C, Aarseth S J and Ferraz-Mello S 1994 
Resonance capture and the formation of the outer planets
\emph{Mon. Not. R. Astron. Soc.} {\bf 270} 21--34
\newline \mhs Melita M D and Woolfson M M 1996 Planetary commensurabilities
driven by accretion and dynamical friction \emph{Mon. Not. R. 
Astron. Soc.} {\bf 280} 854--862
\newline \mhs Winter O C and Murray C D 1997 Resonance and chaos. I.
First-order interior resonances \emph{Astron. Astrophys.}
{\bf 319} 290--304
\newline \mhs Haghighipour N 1999 Dynamical friction and resonance 
trapping in planetary systems 
{\it Mon. Not. R. Astron. Soc.} 
{\bf 304} 185--194
\newline  \mhs Haghighipour N 1999 Resonance lock and planetary 
dynamics {\it Ph.D. Thesis} University of Missouri-Columbia 
\bibitem{NR2} Wardell Z E 1999 Gravitational radiation reaction 
and the three body problem, in preparation   
\bibitem{smk} Kopeikin S M 1998, private communication
\newline \mhs Kopeikin S M 1997 
Millisecond and binary pulsars as nature's frequency 
standards. I. A generalized statistical model of low-frequency timing noise 
{\it Mon. Not. R. Astron. Soc.} {\bf 288} 129--137
\newline \mhs Kopeikin S M 1999 Millisecond and binary pulsars as 
nature's frequency standards. II. 
Effects of low-frequency timing noise on residuals 
and measured parameters  
{\it Mon. Not. R. Astron. Soc.} 
{\bf 305} 563--590
\bibitem{lbec}Bombelli L and Calzetta E  1992 
Chaos around a black hole {\em Class. Quantum Grav.} {\bf 9} 2573--2599
\bibitem{ihs} Stairs I H, Arzoumanian Z, Camilo F, Lyne A G,
Nice  D J, Taylor J H, Thorsett S E and Wolszczan A  1998 Measurement of 
relativistic orbital decay in the PSR B1534 + 12 binary system {\em 
Astrophys. J.} {\bf 505} 352--357
\newline \mhs Taylor J H and Weisberg J M 1989 \emph{Astrophys. J.}
{\bf 345} 434--450
\bibitem{3} Kovalevsky J 1967 
\emph{Introduction to Celestial Mechanics} ({\it Astrophysics and 
Space Science Library} vol 7)  
(New York: Springer) 
\bibitem{4} Sternberg S  1969 {\it Celestial 
Mechanics} vols 1--2 (New York: Benjamin)
\bibitem{fen} Fenichel N 1971 
Persistence and smoothness of invariant manifolds for flows 
{\em Indiana Univ. Math. J.} 
{\bf 21} 193--226
\bibitem{cr} Robinson C 1983 Sustained resonance for a nonlinear system 
with slowly varying coefficients {\em SIAM J.\ Math.\ Anal.} 
{\bf 14} 847--860
\end{thebibliography}
\end{document}